# A comparison of the new exact solutions of the relativistic wave equations of a charged particle propagating in a strong laser field in an underdense plasma

Sándor Varró

Wigner Research Centre for Physics of the Hungarian Academy of Sciences
Institute for Solid State Physics and Optics, 1121 Budapest, Konkoly-Thege Miklós út 29-33. Hungary

**Abstract.** The relativistic wave equations of a charged particle propagating in a classical monochromatic electromagnetic plane wave, in a medium of index of refraction $n_m < 1$, have been studied. In the Dirac case the found exact solutions [21] are expressed in terms of new complex polynomials, and in the Klein-Gordon case they are expressed in terms of Ince polynomials [22]. In each case these solutions form a doubly infinite discrete set, parametrized by quantized momentum components of the charged particle along the polarization vector and along the propagation direction of the electromagnetic radiation (which may be considered as a plasmon wave of arbitrary high amplitude, propagating in an underdense plasma). This solutions describe a high-contrast periodic structure of the particle density on the plasma length scale, and they may have relevance in the study of novel acceleration mechanisms.

**Keywords:** Dirac equation. Klein-Gordon equation. Strong laser field-matter interactions. Volkov states. Novel conceps of acceleration of particles.

**Introduction**

In the theoretical description of fundamental processes taking place in strong laser fields [1] the Gordon-Volkov states – being the exact solutions of the Klein-Gordon [2] and Dirac equation [3] of a charged particle in vacuum – have played a very important role. Immediately after the constructions of the first laser systems, these solutions started to receive a considerable interest, because they describe analytically the interaction with strong laser fields (in vacuum) 'up to infinite order' [4], [5], [6], [7]. Recently there has been a renoved interest in these wave functions, their investigation is undergoing a certain 'renaissance' [8], [9], [10], [11], [12], [13] , [14]. If the field propagates in a medium then the by now studied solutions of the relativistic wave equations can be expressed by the solutions of the corresponding Mathieu or Hill equations [15], which may be used, in the mathematical study of pair creation in a strong laser field [16], [17], [18], and can be relevant for laser acceleration of particles [19], [20]. These solutions are usually higher transcendental functions, which are related to the Mathieu functions, and cannot be expressed in a finite analytic form.

In the present paper, on the basis of our recent works [21], [22]  (referred to as I and II, resp.) we show that there are closed–form exact solutions of the Dirac and Klein-Gordon equations of a charged particle moving in a strong classical laser field in a medium (which may be an underdense plasma ). In each case, these solutions form a doubly infinite discrete set, labeled by two integer quantum numbers, which represent a quantized spectrum of momentum components along the propagation direction and along the polarization of the electromagnetic radiation. Some basic properties of these new states will be discussed, and numerical illustrations and a brief comparison of the Dirac and Klein-Gordon case will be presented [23], [24], and their possible relevance concerning quantum aspects of novel acceleration methods using high-intensity fields in an underdense plasma will also be mentioned.

## 2. Derivation of the scalar wave equations from the second-order Dirac equation

On the basis of our recent works [21], [22], in the present section we summarize the main steps leading to exact closed-form solutions of the relativistic wave equations of a charged particle interacting with a plane EM field in the presence of a medium. We shall also show the similarities and the differences of the spinor and the scalar description.

The Dirac equation of a spinor particle of charge $e$ and of mass $m$ embedded in an external electromagnetic field, characterized by the four-vector potential $A(x)$ has the form[1]

$$[\gamma \cdot \Pi - \kappa]\psi = 0, \quad \Pi \equiv i\partial - \varepsilon A, \quad \varepsilon = e/\hbar c, \quad \kappa = mc/\hbar, \tag{1}$$

where $c$ is the velocity of light in vacuum, $\hbar$ is Planck's constant divided by $2\pi$ and $\Pi$ is the kinetic four-momentum of the particle. We express the bispinor $\psi = (\gamma \cdot \Pi + \kappa)\Psi$ in terms of a new unknown bispinor $\Psi$, which satisfies the second-order Dirac equation,

$$\left[\Pi^2 - \kappa^2 - \frac{1}{2}\varepsilon\sigma \cdot F\right]\Psi = 0, \quad \sigma \cdot F = \sigma_{\mu\nu}F^{\mu\nu}, \quad \sigma_{\mu\nu} = \frac{i}{2}[\gamma_\mu, \gamma_\nu], \quad F^{\mu\nu} = \partial^\mu A^\nu - \partial^\nu A^\nu, \tag{2}$$

where we have introduced the spin tensor $\sigma_{\mu\nu}$ and the electromagnetic field strength tensor $F^{\mu\nu}$. In a medium of index of refraction $n_m$, a general transverse linearly polarized electromagnetic plane wave of wave vector $k$ can be represented by a vector potential

$$A(x) = e_x A_0 \chi(\xi), \quad \xi = k \cdot x, \quad \{k^\mu\} = k_0(1, 0, n_m, 0), \quad \{e_x^\mu\} = (0, 1, 0, 0), \quad A_0 = F_0/k_0, \tag{3}$$

where $F_0$ denotes the amplitude of the electric field strength, and $\chi(\xi)$ is a dimensionless, differentiable function, which is normalized to unity (i.e $\max |\chi(\xi)|=1$). For a monochromatic wave, we have $\chi(\xi) = \cos\xi$, for instance, and $k_0 = \omega_0/c$, with $\omega_0 = 2\pi\nu_0$ being the circular frequency. In this case (3) represents an $x$–polarized wave which propagates in the positive $y$–direction in the medium, and the argument of the cosine function is $\xi = k \cdot x = \omega_0(t - n_m y/c)$.

---

[1] The Minkowski metric tensor $g_{\mu\nu} = g^{\mu\nu}$ has the components $g_{00} = -g_{ii} = 1$ ($i = 1, 2, 3$) and $g_{\mu\nu} = 0$ for $\mu \neq \nu$ ($\mu, \nu = 0, 1, 2, 3$). The scalar product of two four-vectors $a$ and $b$ is $a \cdot b = g_{\mu\nu}a^\mu b^\nu$, i.e. $a \cdot b = a_\nu b^\nu = a^0 b^0 - \boldsymbol{a} \cdot \boldsymbol{b}$, where $\boldsymbol{a} \cdot \boldsymbol{b}$ is the usual scalar product of three-vectors $\boldsymbol{a}$ and $\boldsymbol{b}$. The four-gradient is defined as $\partial = \{\partial^\mu = \partial/\partial x_\mu\} = (\partial/\partial ct, -\partial/\partial \boldsymbol{r})$, where $x = \{x^\mu\} = (ct, \boldsymbol{r})$ is the four-position. In terms of the Dirac matrices $\boldsymbol{\alpha} = (\alpha_x, \alpha_y, \alpha_z)$ and $\beta$, the $\gamma$ matrices are expressed as $\gamma^{1,2,3} = \gamma_{x,y,z} = \beta\alpha_{x,y,z}$ and $\gamma^0 = \beta$, their commutation relations read $\gamma^\mu \gamma^\nu + \gamma^\nu \gamma^\mu = 2g^{\mu\nu}$.

In order to derive the relevant set of scalar equations from (2), we have to solve the algebraic eigenvalue equation of the matrix part of the EM-field-spin interaction term $\sigma \cdot F = 2iA_0 \chi'(\xi)(\gamma \cdot k)(\gamma \cdot e_x)$. This can be simply done [21] in the Majorana representation of the Dirac matrices, yielding the twofold degenerate eigenvalues $\pm \lambda$,

$$(\gamma \cdot k)(\gamma \cdot e_x) u_s = k_0 \lambda_s u_s, \quad \lambda_1 = \lambda_2 = +\sqrt{1 - n_m^2} \equiv \lambda, \quad \lambda_3 = \lambda_4 = -\sqrt{1 - n_m^2} = -\lambda \quad (n_m < 1). \tag{4}$$

Now, if $\Psi$ in Eq. (2) is proportional to one of the eigenvectors $u_s$, then the matrix operation is simply a multiplication with the corresponding eigenvalue $\lambda_s$. In general $\Psi = \sum_{s=1}^{4} \Psi_s u_s$, where $\Psi_s$ satisfies

$$[\Pi^2 - \kappa^2 - i\varepsilon F_0 \chi'(\xi) \lambda_s] \Psi_s = 0 \quad (s = 1-4). \tag{5}$$

By using the ('positive energy') de Broglie plane wave Ansatz and Ince's transformation (Eqs. (5a) and (9a) of I, resp.) in Eq. (5),

$$\Psi_s = \Psi_{ps}(\xi) \exp\left\{-i\left[p - \frac{(k \cdot p)}{k^2} k\right] \cdot x\right\}, \quad \Psi_{ps}(\xi) = f \exp[-(a/4)\cos\xi], \quad a \equiv 4\frac{|\varepsilon| A_0}{\sqrt{k^2}}, \tag{6a}$$

we can derive a complex generalization of Ince's equation (see Eq. (10a) in I for $\varepsilon < 0$, $s = 1,2$),

$$\frac{d^2 f}{dz^2} + a \sin 2z \left(\frac{df}{dz} + if\right) + (\eta - qa\cos 2z) f = 0, \quad z \equiv \xi/2, \quad 2p_x \equiv (q+1)k_p, \quad \eta = 4(k \cdot p)^2 / k_p^4. \tag{6b}$$

By neglecting the spin-EM-field interaction term, Eq. (5) reduces to the Klein-Gordon equation

$$[\Pi^2 - \kappa^2] \Phi = 0. \tag{7}$$

Along an analogous procedure as above (see Eqs. (5a) and (9) of II for $\varepsilon < 0$), we use the Ansätze

$$\Phi = \Phi_p(\xi) \exp\left\{-i\left[p - \frac{(k \cdot p)}{k^2} k\right] \cdot x\right\}, \quad \Phi_p = w \exp[-(a/4)\cos\xi], \quad a = 4|\varepsilon| A_0 / k_p, \quad k_p \equiv \sqrt{k^2}, \tag{8a}$$

and now we arrive at Ince's equation (see Eq. (10a) in II),

$$\frac{d^2 w}{dz^2} + a \sin 2z \frac{dw}{dz} + (\eta - qa\cos 2z) w = 0, \quad z \equiv \xi/2, \quad 2p_x \equiv (q+1)k_p, \quad \eta = 4(k \cdot p)^2 / k_p^4. \tag{8b}$$

In the last equations of both Eqs. (6b) and (8b) we have used the restricted definition on the parameter $\eta$, namely, we restricted the four-momentum $p_\mu$ to satisfy the dressed mass-shell relation $p^2 = \kappa^2 + \varepsilon^2 A_0^2 \equiv \kappa_*^2$, where $\kappa_*$ contains the intensity-dependent mass-shift [4], [14]. Concerning further mathematical details of the derivations, we refer the reader to our recent works [21], [22]. We note that in each cases the modulation functions $\Psi_{ps}$ and $\Phi_p$ satisfy Whittaker-Hill type equations [15], and researches on the asymptotic behaviour of the so-called instability zones of these equations has recently led to significant results [25], which may have importance in applying our solutions.

We also note that throughout the present paper we assume $n_m$ being smaller than unity, in which case the wave vector is time-like ($k^2 > 0$). As a physical example for such a situation we may consider an EM radiation propagating in an underdense plasma, whose dielectric permittivity $\varepsilon_m(\omega) = n_m^2(\omega)$ is taken from the Drude model, i.e. $\varepsilon_m(\omega) = 1 - \omega_p^2/\omega^2$. Here $\omega_p$ is the plasma frequency, with $4\pi n_e e^2/m = \omega_p^2 < \omega^2$, where $n_e$ is the density of a free electron medium, and $k_p \equiv \sqrt{k^2} = \omega_p/c$ does not depend on the frequency. Corresponding to the phase $\xi = \omega(t - n_m y/c)$ of the laser field in the medium, the electromagnetic plane wave under discussion has the dispersion relation $\omega(k_y) = \sqrt{\omega_p^2 + (ck_y)^2}$, with $k_y = n_m\omega/c$. Its phase and group velocities are $v_{ph} \equiv \omega/k_y = c/n_m$ and $v_{gr} \equiv \partial\omega(k_y)/\partial k_y = cn_m$, respectively, which satisfy the general relation $v_{ph}v_{gr} = c^2$, as they should.

## 3. Discrete particle momenta associated to the closed-form basically-periodic solutions of the relativistic wave equations

We have recently shown that if in Eq. (6b) $q + 1 = 2n$ is an even integer, which means that the transverse momentum component $p_x$ is an integer multiple of the 'plasma momentum' $k_p$ (i.e. $p_x = \frac{1}{2}(q+1)k_p = nk_p$), then there are finite-term, polynomial solutions of the second order Dirac equation (see Eqs. (13a-b-c) in I). There are $(2n)$ real linearly independent vectors $\{D_r^{(k)}\}$, associated to the real and different eigenvalues $\eta_n^{(k)}$ $(1 \leq k \leq 2n)$.

$$f = g_n^k(\xi|a,+) = \sum_{r=-n+1}^{n} D_r^{(k)}(a|2n)\exp(-ir\xi) \quad (n = 1, 2, ...), \quad (1 \leq k \leq 2n). \tag{9}$$

Analogously, in the case of a Klein-Gordon particle, Eq. (8) also has several (four) type of finite solutions, if $q$ is an even or odd integer. We have for example shown in II (see Eqs. (11-16) of II), that if $q = 2n$, i.e. if $p_x = k_p(n + \frac{1}{2})$, then the solutions of (8b) reduces to the Ince polynomial $C_{2n}^{2m}(\xi/2, a)$,

$$w = \varphi_n^k(\xi|a,c) = \sum_{r=0}^{n} A_r^{(k)}(a|n+1)\cos r\xi \quad (n = 1, 2, ...), \quad (0 \leq k \leq n). \tag{10}$$

There are $(n+1)$ real linearly independent eigenvectors $\{A_r^{(k)}\}$, associated to the eigenvalues $\eta_n^{(k)}$ $(0 \leq k \leq n)$. These solution may be called 'even solution' of (10a). In the notation $\varphi_n^k(\xi|a,c)$ the letter

"c" refers to the word „cosine", and in $A_r^{(k)}(a|n+1)$ the second argument refers to that there are $n+1$ sets of the coefficients (eigenvectors), and the superscript labels the $k-$th such set.

In the rest of the present section, we present a few figures showing some characteristic properties of the solutions given by Eqs. (9) and (10).

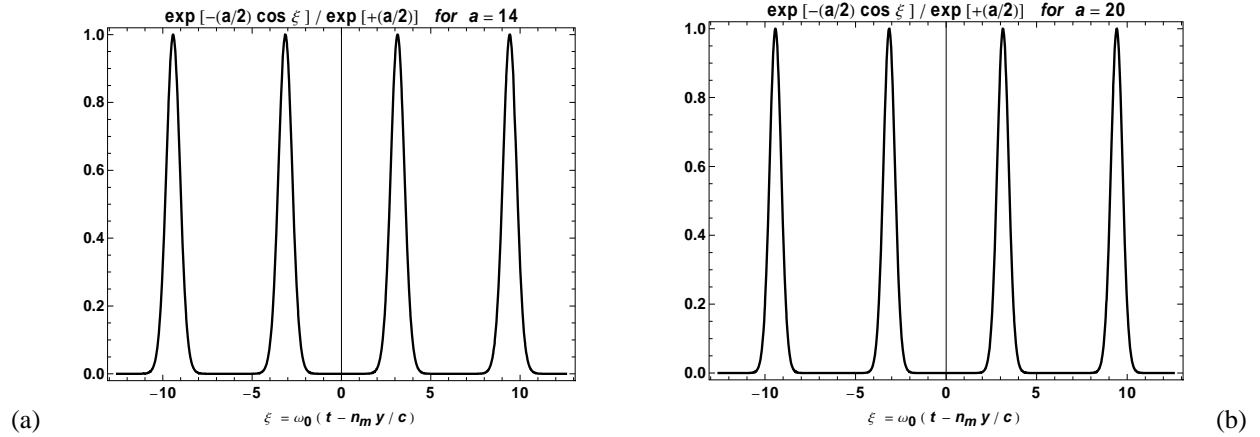

(a)   (b)

**Figure 1.** Shows the normalized square of the envelope function, $\exp[-(a/2)\cos\xi]$, in the wave function (6a) or (8a) for two values of the fundamental parameter $a = 4\,|\,\varepsilon\,|\,A_0/k_p$. (a): $a = 14$, (b): $a = 20$. At the center of the base interval $-\pi \leq \xi \leq +\pi$ (modulo $2\pi$) the 'probability density' is practically zero; this void region represents a sort of 'quantum bubble'.

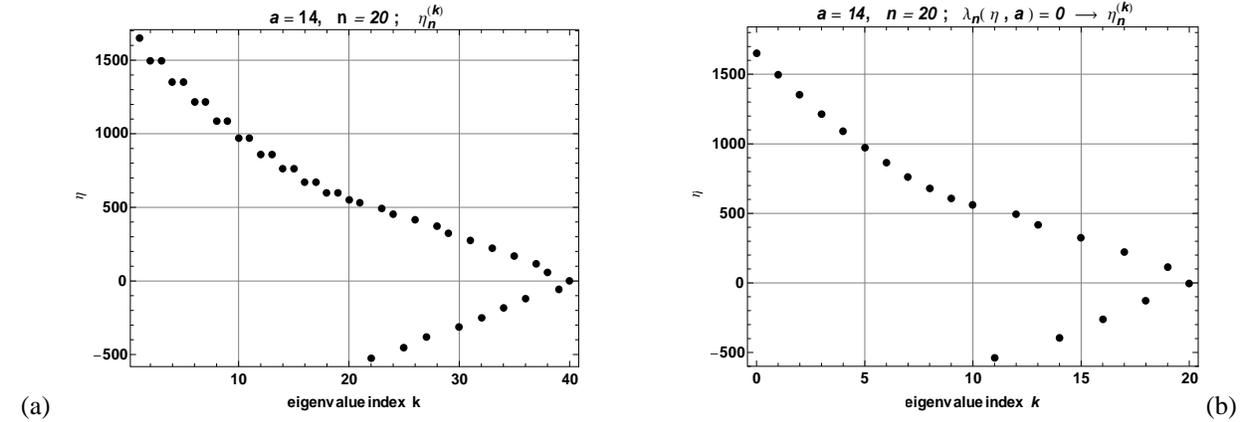

(a)   (b)

**Figure 2.** (a) Shows the eigenvalues associated to the solutions Eq. (9), for $a = 14$ and $n = 20$. We have taken $\hbar\omega_p = 1eV$ and $n = 20$, i.e. the transverse momentum of the electron (in original units) is $p_x = 20 \times \hbar k_p$, where $k_p = \omega_p/c$. The interaction with a Ti:Sa laser field of photon energy $\hbar\omega_0 = 1.563eV$ and peak intensity $I_0 = 100MW/cm^2$ has been considered as an example. (b) shows the same for a Klein-Gordon particle, according to the solution Eq. (10). In this case there are only 20 real eigenvalues, in (a) we see a sort of 'hyperfine splitting' in the Dirac case.

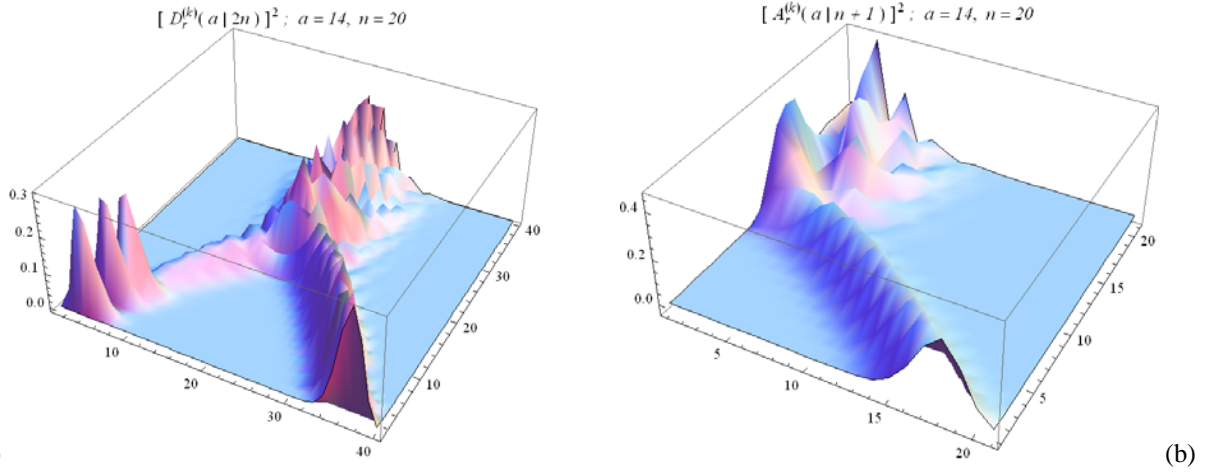

**Figure 3.** Shows an overview of the strength $[D_r^{(k)}(a|2n)]^2$ in Eq. (9) for a Dirac (a), and $[A_r^{(k)}(a|n+1)]^2$ in Eq. (10) for a Klein-Gordon particle of the harmonic coefficients on a three-dimensional list plot when $a=14$ and $n=20$. This figure summarizes the behaviour of these quantities for different eigenvalues (of indeces $k=1, 2, ..., 40$ and $k=1, 2, ..., 20$, which are drawn on the right axis). The discrete points are connected by a smoothened surface in order to guide the eye. (We note that the true $r-$values run from $-20$ to $+20$ in the Dirac case.) For the lowest index $k=1$ the distribution is concentrated to positive $r-$values (left axis) in a single peak, the negative index Fourier coefficients are practically zero in the Klein-Gordon case. For a Dirac particle (a) for larger $k-$values the two peaks merge to a oscillatory distribution.

## 4. Summary

We have presented a brief comparative study of closed-form exact solutions of the Dirac equation and of the Klein-Gordon equation of a charged particle moving in a monochromatic classical plane electromagnetic field in a medium of index of refraction $n_m < 1$. These solutions, labelled by two integer quantum numbers, are finite (complex) trigonometric polynomials in the Dirac case, and they are expressible in terms of the known Ince polynomials in the Klein-Gordon case. As an example for a physical interpretation of these solutions, we have considered the medium an underdense plasma. In this case one of the quantum numbers $n$ characterizing the transverse momentum of the test charge along the polarization, such that either $p_x = nk_p$ or $p_x = (n+½)k_p$, where $k_p = \omega_p/c$, with $\omega_p$ being the plasma frequency. The other quantum number determines the possible values of the longitudinal momentum parameter, $\eta \sim (k \cdot p)^2$. According to our solutions, there exists an interrelation between the quantized transverse and longitudinal momenta, this coupling naturally comes out from the formalism. The existence of the basically-periodic 'bound states' we have found, may have relevance concerning possible quantum features of mechanisms of laser acceleration of electrons by high-intensity laser fields and wake-fields in an underdense plasma [19]. The fundamental parameter $a$ which determines the strength of the interaction is the work done on the charged particle by the electric force of the laser field along the plasma wavelength divided by the photon energy. This $a$ is a quantum parameter, it does not depend on the mass of the particle, and it is typically by many orders of magnitudes larger than the usual intensity parameter $\mu_0$. We have found 'void regions' in the particle density, centered in the middle of the plasma wave cycle with an exponentially large contrast ($e^{a/2}$). This behaviour of the density corresponds to a formation of a perfectly regular and periodic stucture on the plasma length scale, and they may have relevance in the study of novel acceleration mechanisms [26].

**Acknowledgments**

This work has been supported by the Hungarian Scientific Research Foundation OTKA, Grant No. K 104260.